\documentclass[12pt]{iopart}
\usepackage{graphicx}
\usepackage{psfig}

\begin{document}

\title[High $p_T$ particles in d+Au and Au+Au at PHENIX]{PHENIX measurement of high $p_T$ particles in Au+Au
and d+Au collisions at $\sqrt{s_{NN}} = 200$~GeV}

\author{Christian Klein-B\"osing\dag\ for the PHENIX Collaboration
\footnote[3]{For the full PHENIX Collaboration author list and
acknowledgments, see Appendix "Collaborations" of this volume.}  }

\address{\dag\ Institut f\"ur Kernphysik, Wilhelm-Klemm-Str. 9, D-48149 M\"unster, Germany}

\begin{abstract}

We present new results on the production of charged hadrons and
neutral pions in Au+Au and d+Au collisions at $\sqrt{s_{NN}} =
200$~GeV, measured within the PHENIX experiment. We observe the opposite
centrality dependence when comparing the two collision systems: an
increase with centrality compared to the binary scaled yield 
from p+p collisions in d+Au and a decrease with centrality in Au+Au.

\end{abstract}




\section{Introduction}


The production of hadrons with large transverse momentum ($p_{T}$) is
dominated by the fragmentation of quarks and gluons, produced in
parton-parton scatterings with large momentum transfer $Q^2$. The
production cross section for this hard scattering depends on the
initial distribution of partons in the colliding species, the
elementary parton-parton cross section and the fragmentation process
of partons into hadrons.

In absence of any medium effects the production of high $p_T$
particles in Au+Au and d+Au collisions should be comparable to the
production in p+p after scaling with a geometrical factor which reflects
the increased number of scattering centers. The geometrical factor
for a given centrality is usually expressed in terms of the {\sl
nuclear overlap function} $T_{AB}$ or the number of binary
nucleon-nucleon collisions $N_{coll}$. The comparison of different
colliding species or centralities to the p+p reference is done with
the {\sl nuclear modification factor}:

\begin{equation}
R_{AB} = \frac{\rmd^2N_{AB}/ \rmd y \rmd p_{T}}
{T_{AB} \cdot \rmd^2 \sigma_{pp}/ \rmd y \rmd p_{T}}.
\end{equation}

If the model of geometrical scaling from p+p to other colliding
species is valid, the nuclear modification factor should be unity
above a certain $p_T$ where hard scattering is dominant. In contrast
to this expectation the results for Au+Au collisions at the
Relativistic Heavy Ion Collider (RHIC) showed a suppression of up to a
factor of five in the nuclear modification factor
\cite{Adler:2003qi,Adler:2003au}.

It has been suggested that the observed suppression is a result of
parton energy loss in a dense medium like the quark-gluon plasma
\cite{Gyulassy:1990ye,Wang:2003yp}. However, with only Au+Au collisions it is
hardly possible to distinguish between final state effects and effects
of cold nuclear matter at $\sqrt{s_{NN}} = 200\,$GeV (initial state
effects).

Initial state effects can influence the particle production in various
ways. A modification of the parton distribution function in a
nucleus can result in a suppression or enhancement of particle
production, depending on the momentum fraction $x$ of the scattered
parton ({\sl shadowing} or {\sl anti-shadowing}). Multiple soft
scattering of incoming nucleons or partons leads to an enhancement at
intermediate $p_{T}$ ({\sl Cronin Effect}). Such effects are also
present in d+Au collisions, while no large volumes with increased
energy density are formed. These reactions have been studied in early
2003 at the Relativistic Heavy Ion Collider (RHIC) and first results
have been published in \cite{Adler:2003ii}, showing the absence of
suppression in minimum bias reactions at mid-rapidity.

The examination of d+Au collisions also tests whether more exotic
phenomena, e.g. the possible formation of a {\sl Color Glass
Condensate} (CGC) in the initial state at $\sqrt{s_{NN}} = 200$~GeV
can be responsible for the observed suppression in central Au+Au
collisions. In this case the nuclear modification factor in d+Au
should also show a decrease with centrality and a $R_{AB}$ below unity
in central d+Au collisions \cite{Kharzeev:2002pc}.

\section{Data analysis and results}

The data we present were collected in two different RHIC runs. In
2001--2002 Au+Au collisions at $\sqrt{s_{NN}} = 200$~GeV and in early
2003 d+Au collisions at the same energy were studied. 
The PHENIX central arm detectors used in this analysis provide a solid
angle coverage of $\Delta\eta = 0.7$ and $\Delta\phi = \pi$. They have the
capability of identifying neutral pions and charged hadrons over a
broad momentum range \cite{Aphecetche:zr,Adcox:zp}.

\begin{figure}
  \begin{minipage}{7.5cm}
   \hspace{2cm}\includegraphics[width=5.5cm]{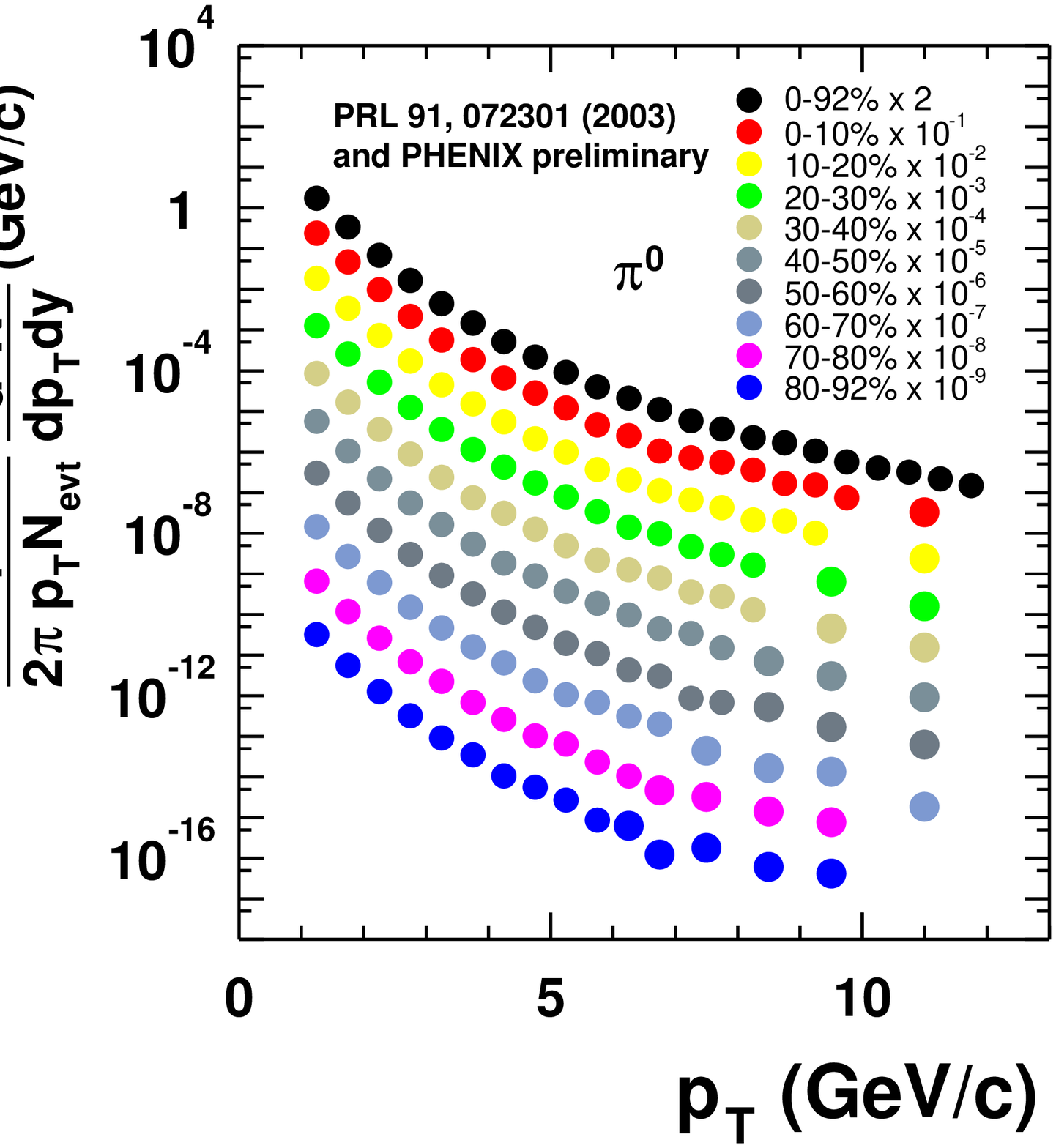}
   \caption{Production of neutral pions in Au+Au collisions for
   different centralities. 
}
   \label{fig:pro_pi0_spectra_AuAu}
   \end{minipage}
  \begin{minipage}{7.5cm}
   \hspace{2cm}\includegraphics[width=5.5cm]{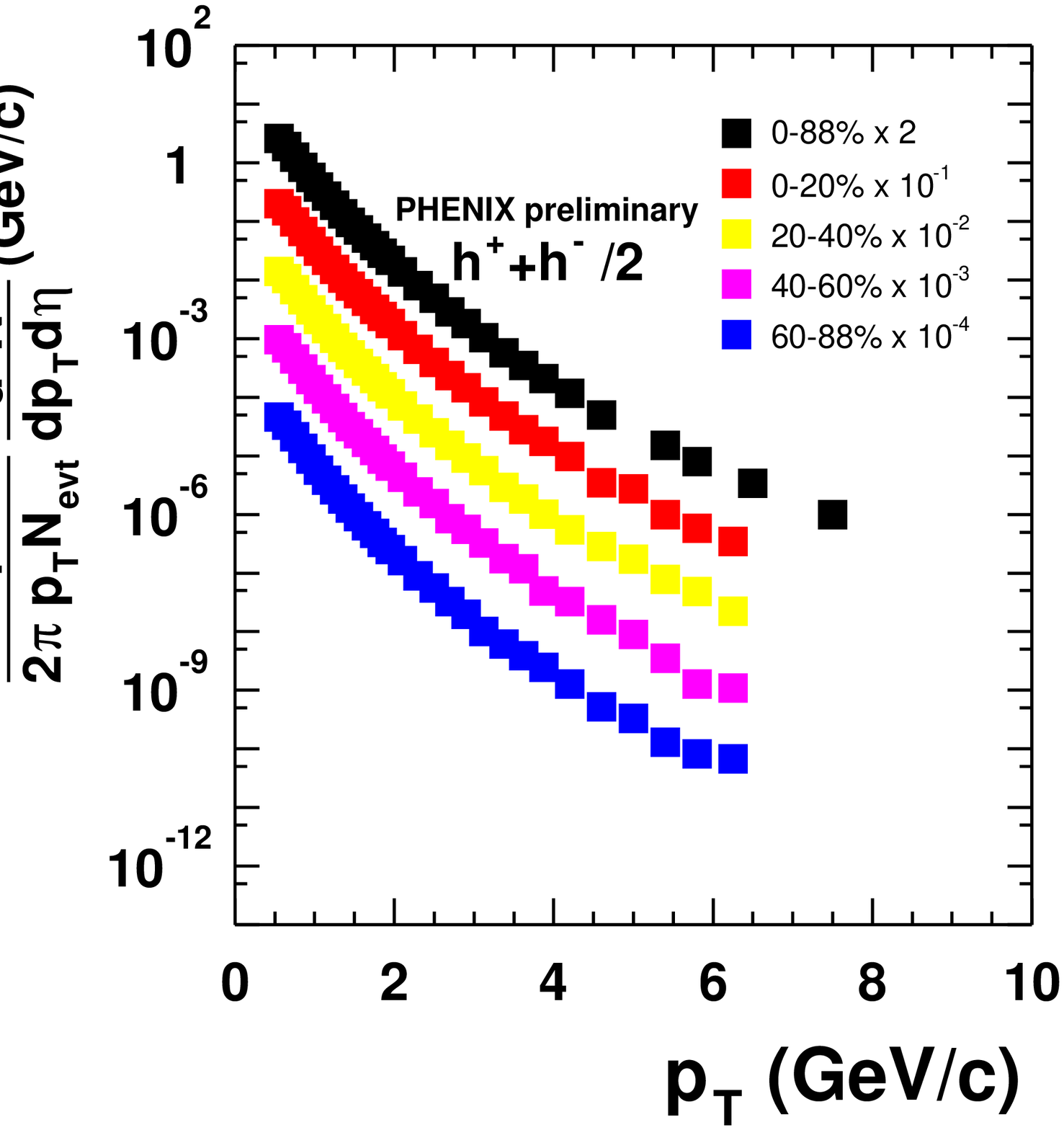}
   \caption{Production of charged particles in d+Au collisions for
   different centralities.} 
   \label{fig:pro_charged_spectra_dAu}
   \end{minipage}
\end{figure}
\begin{figure}[!t]
   \centerline{\includegraphics[height=15.5cm]{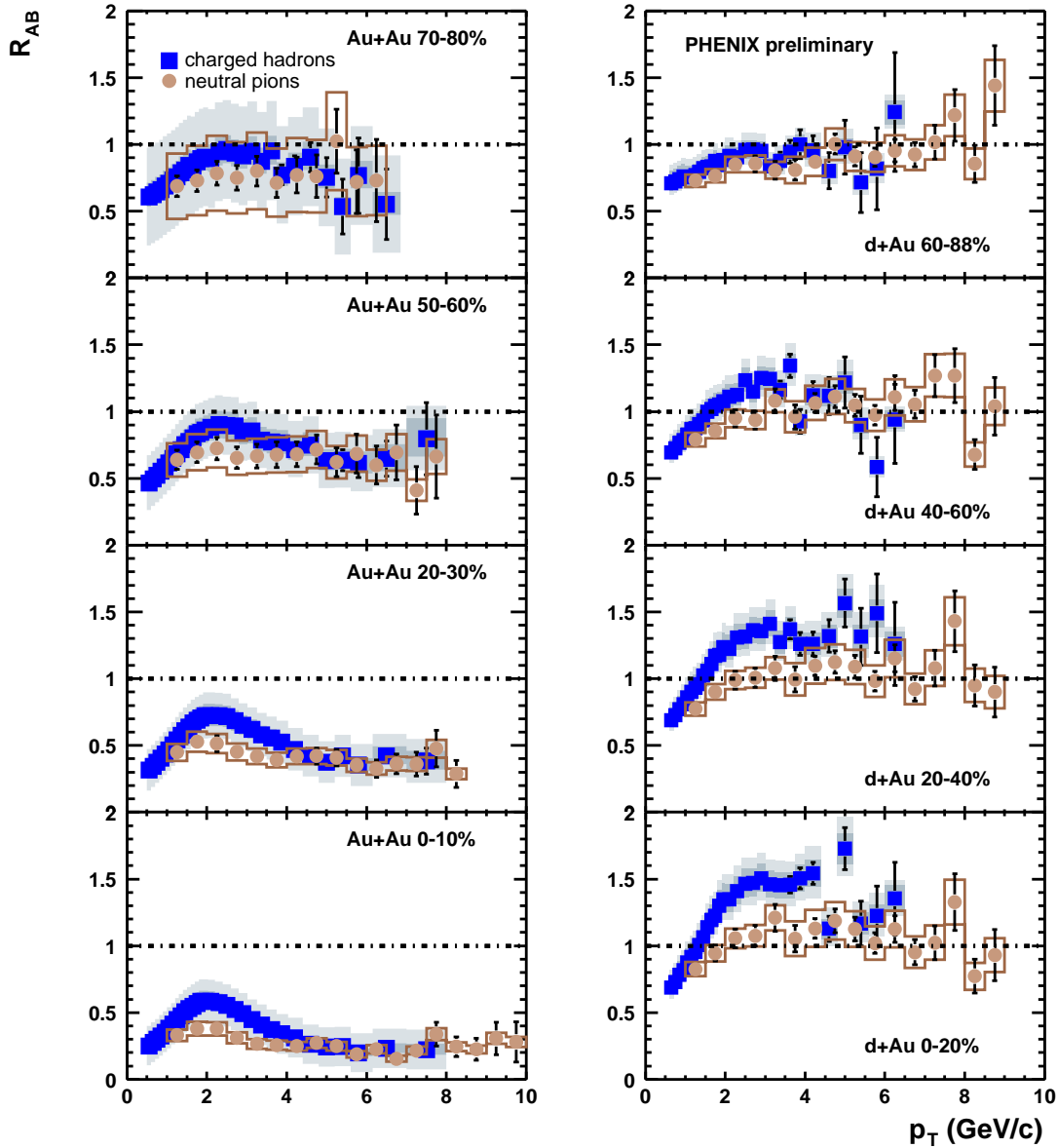}}
   \caption{Centrality dependence of the nuclear modification factor
   in Au+Au (left) and d+Au (right)  collisions.}
   \label{fig:pro_cent_dep}
\end{figure}

Neutral pions are reconstructed via their 2$\gamma$ decay channel,
using an invariant mass analysis of the photon pairs measured in the
PHENIX Electromagnetic Calorimeter (EMCal). The combinatorial
background of the invariant mass analysis is determined by a {\sl
mixed event} technique using photon pairs from different events
but similar multiplicity and vertex position.  The raw $\pi^0$ spectra
are corrected for geometrical acceptance and multiplicity dependent
reconstruction and efficiency losses.

The data set used for the $\pi^0$ analysis in Au+Au consists of 30~M
minimum bias events and a newly incorporated data set of
high-energy photon triggered events which corresponds to an additional
50~M minimum-bias events. The enlarged data set extends the $p_{T}$
reach up to 10~GeV/$c$ (see Fig.~\ref{fig:pro_pi0_spectra_AuAu}) for
all centralities compared to the result in \cite{Adler:2003qi}.
In the analysis of the d+Au run 16~M minimum-bias events have been
combined with an equivalent of 100~M minimum-bias events from a
high-$p_{T}$ photon triggered data set, resulting in the measurement
of $\pi^{0}$'s up to 10~GeV/$c$.

Charged particles are tracked through the PHENIX central arms by Pad
Chambers and Drift Chambers. Particle momenta are reconstructed by
their deflection in an azimuthaly symmetric magnetic field with a
precision of $\delta p/p \simeq 1 \% \oplus 1 \% \cdot p$/(GeV/$c$)
using tight matching criteria in the Pad Chambers to reduce the
background. The data sets analyzed were 27~M minimum-bias events in
Au+Au and 12~M events in d+Au, which allows to study unidentified charged
particles in a broad momentum range (See
Fig.~\ref{fig:pro_charged_spectra_dAu} and \cite{Adler:2003au}).


To calculate the nuclear modification factor the PHENIX p+p $\pi^0$
spectrum is used \cite{Adler:2003pb}. In case of the charged particles
these data were scaled by a factor of 1.6 to account for the known
abundance of total charged particles over $\pi^0$'s and combined with
the data from UA1 for charged hadrons \cite{Albajar:1989an}. The
result for different centralities is shown in Fig.
\ref{fig:pro_cent_dep} for Au+Au and d+Au. In very peripheral
collisions charged particle and neutral pion yields are consistent
with the expectation from binary scaling in both colliding
systems. Going to more central collisions the suppression gets stronger
in Au+Au, in contrast to the situation in d+Au where the particle
production is enhanced compared to the expectation from binary scaled
p+p. Another interesting observation in central collisions is the
difference between charged hadrons and neutral pions in both colliding
systems. This is mainly due to an anomalous p/$\pi$ ratio at
intermediate $p_{T}$. For further discussion see also \cite{felix}.
 
\section{Conclusion}

We have presented results on the centrality dependent production of
neutral pions and charged hadrons in d+Au and Au+Au collisions at
$\sqrt{s_{NN}} = 200$~GeV. The d+Au measurement provides the baseline
for effects of cold nuclear matter at mid-rapidity for this energy.
The absence of suppression in d+Au collisions shows that the suppression
observed in central Au+Au collisions cannot be attributed to
initial state effects, like shadowing or the formation of a Color
Glass Condensate. The observation supports the picture of a
medium induced energy loss.


\begin{thebibliography}{99}

\bibitem{Adler:2003qi}
PHENIX Collaboration 
Adler~S~S {\it et al.} 2003  
{\sl Phys.\ Rev.\ Lett.\ }  {\bf 91} 072301 

\bibitem{Adler:2003au}
PHENIX Collaboration 
Adler~S~S {\it et al.} 2003 
arXiv:nucl-ex/0308006 {\sl to appear in Phys. Rev.} C

\bibitem{Gyulassy:1990ye}
Gyulassy~M and Plumer~M 1990
{\sl Phys.\ Lett.\ } B {\bf 243} 432 


\bibitem{Wang:2003yp}
Wang~X~N 2003
{\sl Nucl.\ Phys.} A {\bf 715} 775 

\bibitem{Adler:2003ii}
PHENIX Collaboration 
Adler~S~S {\it et al.} 2003
{\sl Phys.\ Rev.\ Lett.\ }  {\bf 91} 072303 


\bibitem{Kharzeev:2002pc}
Kharzeev~D, Levin~E and McLerran~L 2003
{\sl Phys.\ Lett.} B {\bf 561} 93 


\bibitem{Aphecetche:zr}
PHENIX Collaboration
Aphecetche~L {\it et al.} 2003 
{\sl Nucl.\ Instrum.\ Meth.\ } A {\bf 499} 521 


\bibitem{Adcox:zp}
PHENIX Collaboration
Adcox~K {\it et al.} 2003  
{\sl Nucl.\ Instrum.\ Meth.\ } A {\bf 499} 489 


\bibitem{Adler:2003pb}
PHENIX Collaboration
Adler~S~S {\it et al.} 2003
{\sl Phys.\ Rev.\ Lett.\ }  {\bf 91} 241803 

\bibitem{Albajar:1989an}
UA1 Collaboration
Albajar~C {\it et al.} 1990
%
{\sl Nucl.\ Phys.\ } B {\bf 335} 261 

\bibitem{felix} Matathias F for the PHENIX Collaboration,
                   these proceedings.






\end{thebibliography}
\end{document}